\documentclass[
]{ceurart}

\usepackage{listings}
\usepackage{xcolor}
\usepackage[acronym]{glossaries}
\newacronym{ml}{ML}{Machine Learning}
\newacronym{se}{SE}{Software Engineering}

\begin{document}

\copyrightyear{2025}
\copyrightclause{Copyright for this paper by its authors.
  Use permitted under Creative Commons License Attribution 4.0
  International (CC BY 4.0).}

\conference{IWST 2025: International Workshop on Smalltalk Technologies, July 1-4, 2025, Gdansk, Poland}

\title{Analysing Python Machine Learning Notebooks with Moose}

\author[1]{Marius Mignard}[]
\author[1]{Steven Costiou}[]
\author[1]{Nicolas Anquetil}[]
\author[1]{Anne Etien}[]

\address[1]{Univ. Lille, CNRS, Inria, Centrale Lille, UMR 9189 CRIStAL, F-59000 Lille, France}

\begin{abstract}
Machine Learning (ML) code, particularly within notebooks, often exhibits lower quality compared to traditional software.
Bad practices arise at three distinct levels: general Python coding conventions, the organizational structure of the notebook itself, and ML-specific aspects such as reproducibility and correct API usage.
However, existing analysis tools typically focus on only one of these levels and struggle to capture ML-specific semantics, limiting their ability to detect issues.
This paper introduces Vespucci Linter, a static analysis tool with multi-level capabilities, built on Moose and designed to address this challenge.
Leveraging a metamodeling approach that unifies the notebook's structural elements with Python code entities, our linter enables a more contextualized analysis to identify issues across all three levels.
We implemented 22 linting rules derived from the literature and applied our tool to a corpus of 5,000 notebooks from the Kaggle platform.
The results reveal violations at all levels, validating the relevance of our multi-level approach and demonstrating Vespucci Linter’s potential to improve the quality and reliability of ML development in notebook environments.
\end{abstract}


\maketitle

\section{Introduction}

Despite the exponential growth of software development in \gls{ml} and related fields, such as data science, recent studies indicate that \gls{ml} code often shows lower quality than traditional software \cite{10.1145/3382494.3410680, 10356405}. One reason is that many \gls{ml} developers come from non-software backgrounds, making them unfamiliar with established best practices in \gls{se} or prone to adopting alternative coding conventions \cite{10.1145/3382494.3410680}. Additionally, much of \gls{ml} development follows the literate computing paradigm \cite{10.1093/comjnl/27.2.97, stodden2014implementing}, which combines text, code, and outputs—most often in notebooks such as \textit{Jupyter Notebook}\footnote{\url{https://jupyter.org/}} \cite{notebook}.
While notebooks facilitate rapid prototyping, ease of documentation, and sharing, they often lack rigorous enforcement of coding standards due to limited integration with code quality tools such as linters or code-smell detectors.

Previous studies have worked on formalizing coding rules, smells, and bad practices in both notebooks and \gls{ml} applications \cite{7780188, 10.1145/3512934, 10.1145/3522664.3528620}. These practices can be divided into three levels. The first is the Python level, which involves following PEP8-style\footnote{\url{https://peps.python.org/pep-0008/}} rules, creating well-structured functions, and writing tests. The second is the notebook level, such as including \textit{Markdown} cells to explain the code, placing imports at the beginning, and recording library versions. Finally, there are \gls{ml}-specific rules, like setting random seeds for reproducibility and using recommended methods with the correct parameters. 

Single-level linters work well for conventional Python projects, but ML notebooks introduce challenges that current tools struggle with: fragmented context across cells and ML-specific semantics that go beyond style or AST-only checks and often span multiple cells. Vespucci Linter addresses these challenges by unifying notebook structure and Python code entities into a single model, so rules can be expressed at the Python, notebook, and ML levels within one analysis. While most current rules are single-level, the unified model already supports cross-level reasoning. 

The notebook quality issue is critical as notebooks increasingly serve as vehicles for sharing analytical insights, prototypes, and educational materials \cite{10.1145/3173574.3173606, 10.1145/3377816.3381724}, often targeting audiences with limited programming expertise (and prone to reuse code \cite{9127202}). Without tools capable of enforcing good practices across all three rule levels, poor coding habits are propagated, potentially resulting in production-level systems built on suboptimal \gls{ml} models \cite{SE_AI}. This can create a negative feedback loop, perpetuating poor quality code and models\footnote{\label{fn:model-meaning}It is important to note that here, we are referring to ML models that enable prediction. In the remainder of the text, the term *model* refers to its software engineering meaning—an abstract representation of a system conforming to a metamodel.} through successive generations of learners and developers.

Although various analysis tools are available, including some dedicated to data science (data validation or correct use of data-related functions) \cite{10.1145/3689609.3689996}, only one tool has recently started to incorporate aspects of the machine learning \emph{workflow}.
\noindent
This highlights the need for specialized analysis tools tailored to notebook-based \gls{ml} development, where code is fragmented across multiple cells interleaved with text and outputs. Furthermore, such tools must incorporate \gls{ml}-specific semantic analysis, addressing issues like appropriate method calls, correct parameter usage, and proper sequencing of operations. Existing tools typically focus either on general Python coding conventions, notebook-level structural recommendations, or data-level validations, often neglecting the deeper semantic analysis related to the machine learning workflow itself \cite{10.1145/3377816.3381724, 10.1145/3382494.3410680, 10.1145/3689609.3689996}. 

Detecting issues at this ML-specific level poses significant challenges, as it necessitates extracting and interpreting semantic information from Python code to accurately identify the intended logic of \gls{ml} workflows. Misuse or misconfiguration at this level can significantly alter the meaning and validity of the results, thus requiring sophisticated domain-specific analysis.

In response to these challenges, this paper introduces \textit{Vespucci linter}\footnote{\url{https://github.com/JMLF/FamixNotebook}}, a tool built upon the \textit{Moose}\footnote{\url{https://modularmoose.org/}} software analysis platform \cite{10.1007/978-3-030-64694-3_8}. Our tool leverages Moose's modeling capabilities to detect bad practices in \gls{ml} notebooks across all three defined levels. \textit{Vespucci linter} aims to promote best practices within the literate computing paradigm.

This paper is structured as follows. Section 2 reviews related work. Section 3 describes our tool and methodological approach. Section 4 illustrates the application of \textit{Vespucci linter} through a real-world dataset sourced from the Kaggle\footnote{\url{https://www.kaggle.com/}} platform. Finally, Section 5 concludes the paper and outlines potential directions for future research.

\section{Related Work}

\subsection{Python, Notebook And Machine Learning Code Smells}
Code smells are recognized as bad practices, poor design choices or anti-patterns that negatively impact software quality, readability, and maintainability. Unlike bugs, code smells do not necessarily cause direct faults but can lead to other negative consequences \cite{LACERDA2020110610}. Empirical research consistently highlights the negative impacts of code smells, linking them to reduced code quality, higher defect rates, and increased long-term maintenance challenges \cite{10.1145/3180155.3182532, 6392174}. 
Several studies have identified and formalized code smells across the three levels mentioned in the introduction (Python, Notebook, and \gls{ml}).

At the Python level, many guidelines and best practices addressing code smells have already been formalized. A well-known example is the widely adopted linter \textit{PyLint}, which enforces a set of coding conventions derived from established \gls{se} standards. Typical rules include naming conventions, proper structuring of modules and functions, PEP8 style, and the identification of unused variables \cite{pylint_messages}.

Previous works have analyzed Python-based \gls{ml} code using \textit{PyLint} \cite{9474395, 10356405}, and identified that the most common rule violations are related to imports, excessive numbers of function arguments or local variables, pointless statements, and unused variables. 

At the notebook level, studies have proposed specific recommendations to improve notebook usage. For instance, Quaranta et al.\ identified 17 best practices through a systematic literature review \cite{QuarantaA2022}, emphasizing practices such as placing imports at the beginning, using clear notebook names, and including Markdown cells for documentation. Similarly, Rule et al.\ outlined 10 rules to effectively share computational analyses in notebooks, highlighting the importance of splitting cells logically, explicitly recording dependencies, and providing thorough documentation \cite{10.1371/journal.pcbi.1007007}.

At the ML-specific level, Nikanjam et al.\ presented a catalog of 8 design smells, identified through a review of related literature and an analysis of 659 deep learning programs. These smells highlight poor architectural or configuration decisions that negatively impact model performance and maintainability \cite{9609188}. Complementarily, Zhang et al.\ collected 22 ML-specific code smells from academic literature, online forums, and open-source repositories, and proposed corresponding solutions. Their work underlines recurrent issues in \gls{ml} application development, such as uncontrolled randomness and improper hyperparameter configurations \cite{10.1145/3522664.3528620}.

However, existing research mainly addresses code smells independently at one level—either Python, notebook, or \gls{ml}-specific—without considering how these issues interact or compound across multiple levels. Thus, a significant gap remains in understanding the cumulative effects of code smells when combined across different abstraction levels, how smells identified at one level can propagate and impact others and how to globally address them.

\subsection{Existing Quality Control Tools}
Quality control tools currently available primarily cover the Python and notebook-specific levels, along with data validation in data science, but lack dedicated analysis capabilities on \gls{ml} workflows.

At the Python level, widely adopted tools include \textit{PyLint}\footnote{\url{https://www.pylint.org/}}, \textit{PyFlakes}\footnote{\url{https://github.com/PyCQA/pyflakes}}, and \textit{Ruff}\footnote{\url{https://docs.astral.sh/ruff/}}. These tools typically operate by parsing Python code into an Abstract Syntax Tree (AST) and analyzing it against predefined rules. They primarily address general coding issues.

On the notebook level, specialized tools have been proposed. \textit{Pynblint}\footnote{\url{https://github.com/collab-uniba/pynblint}}, developed by Quaranta et al., implements best practices specifically identified for notebooks. It analyzes notebooks based on criteria like proper cell structure, correct usage of Markdown for documentation, and dependency management, providing feedback for improving notebook quality \cite{10.1145/3522664.3528612}. Similarly, \textit{NBLyzer}\footnote{\url{https://github.com/microsoft/NBLyzer}}, proposed by Subotić et al., incorporates the unique cell execution semantics—the possibility of out-of-order cell execution—of notebooks into its analysis. It provides warnings and recommendations on cell ordering, dependencies, and potential data leakage due to improper execution order \cite{10.1145/3510457.3513032}.

Specialized tools also exist for data validation and correct data handling in data science workflows, such as \textit{Pandera}\footnote{\url{https://pandera.readthedocs.io/}} and the prototype linter proposed by Dolcetti et al. These tools specifically target data-related logic errors, ensuring the correctness of data manipulation and transformation steps \cite{10.1145/3689609.3689996}.

In addition, a dedicated tool named \textit{ML Lint} was proposed by van Oort et al. \cite{10.1145/3510457.3513041}. It introduces the novel concept of \emph{project smells} aiming for a more holistic analysis of \gls{ml} projects. \textit{ML Lint} performs static analysis on Python \gls{ml} projects, reviewing source code, data, and configurations. Its linting rules cover five major categories: version control (code and data), dependency management, continuous integration, code quality, and testing practices. \textit{ML Lint} checks for the appropriate use of tools like Git and Data Version Control (DVC), proper separation of development and runtime dependencies, presence of tests, and produces reports highlighting detected project smells. For code issues, \textit{ML Lint} relies entirely on existing Python linters and does not define or enforce any notebook-specific or \gls{ml}–oriented rules.

Another relevant tool is \textit{SonarQube}\footnote{\url{https://www.sonarsource.com/products/sonarqube/}}, a static analysis platform originally developed for general-purpose software quality assurance. Recently, it has started to incorporate rules and practices related to the \gls{ml}-specific level. However, these analyses remain primarily focused on traditional software engineering aspects and do not fully extend to the structure or semantics of \gls{ml} workflows. Moreover, defining additional, domain-specific rules is not straightforward, and the tool does not support analysis at the notebook level.

Existing tools remain mostly focused on single-level analyses, addressing either general Python coding conventions, notebook structure, or data validation. While some tools, such as \textit{SonarQube} have begun to incorporate elements related to \gls{ml}, they do not yet fully address the \gls{ml} workflows-level. 
This gap is problematic, as \gls{ml} workflows involve domain-specific patterns that are not captured by existing tools. There is a need for solutions that extend quality analysis to include the \gls{ml} workflows-level alongside existing checks.

\section{Vespucci Linter}

We designed Vespucci to detect best-practice violations at the \emph{Python}, \emph{notebook}, and \emph{machine-learning} levels within a single analysis tool.
It is based on top of the \emph{Moose} analysis platform, which offers a flexible meta-platform centered around the \emph{Famix} meta-modeling framework. \emph{Famix} provides basic modeling elements that can be combined to define custom metamodels \cite{10.1007/978-3-030-64694-3_8}. Leveraging this flexibility allows us to define and interconnect metamodels: one for Python source code and another for notebooks. Additionally, \emph{Moose} gives us a dedicated model query language and the \emph{Critics}\footnote{\url{https://modularmoose.org/users/moose-ide/moose-critics/}} rule engine, enabling a rapid implementation of new quality rules with just a few lines of Pharo code.  It also allows users to benefit from the broader Moose ecosystem — including existing filtering and grouping functionalities, and visualization tools. Detected violations can be tagged at the level of a cell, function, or any model entity, and then explored visually to reveal structural hotspots.

\subsection{Notebook Meta-model}
\label{sec:metamodel}
Existing linters typically operate on Python's Abstract Syntax Trees (AST). Relying solely on AST requires transforming notebooks into plain Python scripts, discarding positional context information specific to notebook structures, such as cell boundaries. Consequently, these linters lose the granularity required for precise, notebook-specific quality analysis. Similarly, when analyzing notebooks from their JSON structure on a per-cell basis, dependencies across cells may be lost, as individual cells might refer to definitions located in preceding ones that are not part of the current analysis scope.

\begin{figure}[h]
  \centering
  \includegraphics[width=\linewidth]{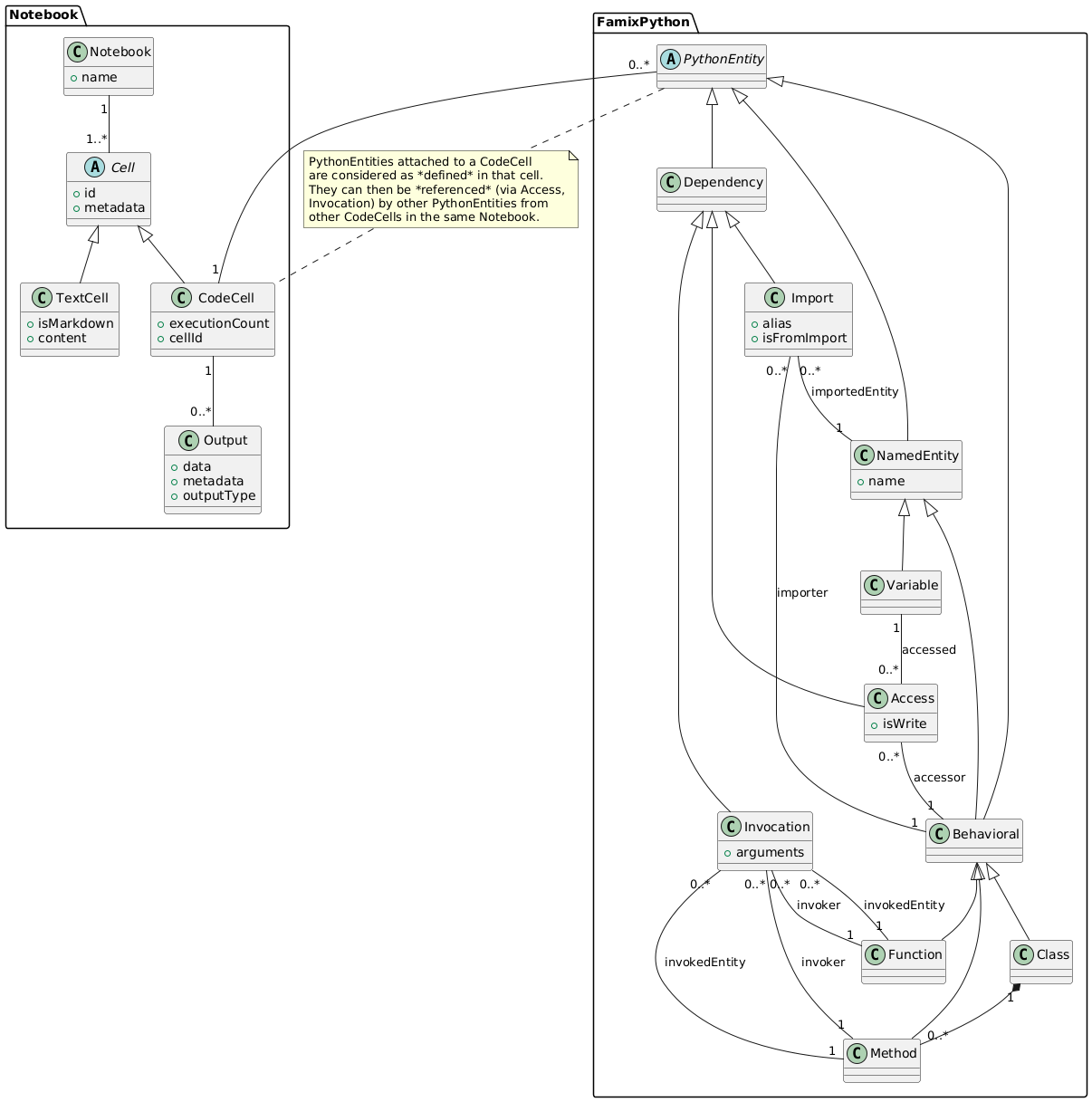}
  \caption{Simplified representation of our notebook meta-model.}
  \label{fig:metamodel}
  \end{figure}

To address these limitations, we propose a \emph{metamodel} (Figure~\ref{fig:metamodel}) to combine notebook structural information with Python elements provided by \textit{FamixPython}. 
Both Figure~\ref{fig:metamodel} and the accompanying explanation present an abstracted view, focusing on the key entities and relationships relevant to our analysis, while deliberately omitting less central elements present in the full \textit{FamixPython} metamodel.

Our metamodel preserves the mapping between Python entities and notebook cells, ensuring that no contextual information is lost.
In our metamodels, a \textit{Notebook} consists of multiple \textit{Cells}, categorized into \textit{CodeCells} and \textit{TextCells}. \textit{CodeCells} hold executable Python code, whereas \textit{TextCells} typically provide markdown documentation. Each \textit{CodeCell} can yield multiple \textit{Outputs}, capturing results from execution, visual outputs, and textual outputs, preserving the full execution context.

The Python aspect of our metamodel leverages the \textit{FamixPython} entities. Each \textit{CodeCell} is directly linked to associated Python entities, including variables, functions, classes, imports, and invocations. This linkage ensures a detailed mapping, enabling queries to precisely trace Python entities back to the cells where they were defined.
Our Python metamodel includes entities containing executable instructions, such as \textit{Behavioral} entities (functions, methods, and classes), and \textit{Association} relationships linking entities. These relationships include \textit{Access} entities capturing read/write operations on variables, \textit{Invocation} entities connecting the invoker to the invoked entity and \textit{Import}. 

This granularity does not stop at cell boundaries but instead provides a notebook-wide view, allowing analyses to span multiple cells, accurately track dependencies, and identify potentially problematic patterns across the entire notebook. By combining these two metamodels, our approach offers rich querying capabilities such as identifying Python imports that do not appear in the first \textit{CodeCell}, or detecting potentially problematic invocation sequences across different notebook cells.

\subsection{Process Overview}
\begin{figure}[ht]
\centering
\includegraphics[width=\linewidth]{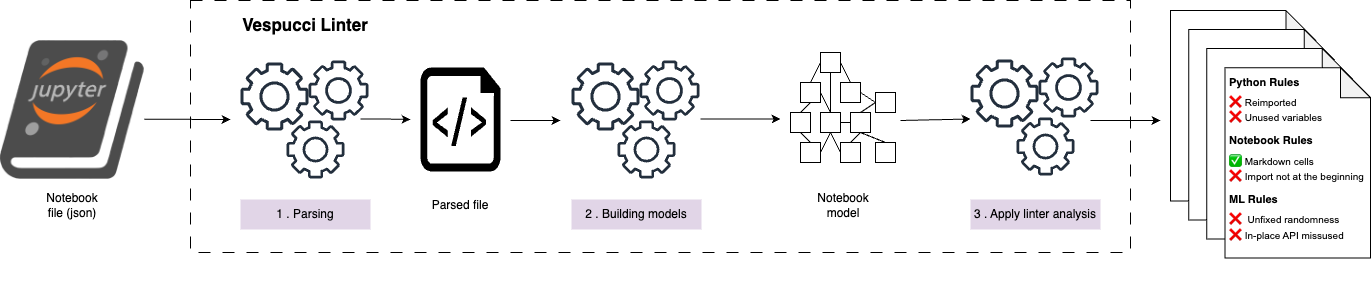}
\caption{Overview of our process to detect rule violations in notebooks.}
\label{fig:method_pipeline}
\end{figure}

The implementation of \textit{Vespucci Linter} follows a four-step pipeline (summarised in Figure~\ref{fig:method_pipeline}) that transforms raw Jupyter notebooks into analyzable models which enable quality analysis without requiring prior code extraction or manual preprocessing. It applies a set of predefined linting rules and reports any detected violations. The following steps describe our approach, from notebook parsing to rule evaluation and result export

\begin{enumerate}
    \item \textbf{Notebook parsing.} 
          The notebook (\texttt{*.ipynb}) file is parsed; cells are extracted while metadata (execution counter, id, type) are preserved.
          We then append all Python code from code cell in a temporary file.
    
    \item \textbf{Notebook modelling.}  
          A \emph{model} is automatically built that merges the \textit{FamixPython} model (from the parse of the previously built temporary file)—capturing imports, functions, invocations\dots—with  
          a custom \textit{Notebook} model—representing \textit{Markdown} and code cells, execution order, and raw metadata.  
    
    \item \textbf{Lint-rule application.}  
          On the generated \emph{model} we evaluate all defined rules.  
          Rules are detailed in Section~\ref{sec:rules}. Each rule returns a \textit{Violation} object with severity, precise location (file/cell), and a suggestion for remediation.
    
    \item \textbf{Result export.}  
          All violations are serialized to JSON for analysis or manual inspection.

\end{enumerate}

\subsection{Proposed Multi-level Rules}
\label{sec:rules}
We selected 22 specific linting rules tailored to address frequent bad practices observed in Python notebooks used for machine learning. These rules originate from the literature and are organized into general Python practices, notebook-specific guidelines, and ML-specific usage patterns. The thresholds and conventions for these rules are extracted from literature but can easily be adapted to specific contexts and libraries.
For clarity, we assign an identifier to each rule, consisting of a letter indicating its level—\emph{P} for Python-level, \emph{N} for notebook-level, and \emph{M} for ML-specific—followed by a number. For example, the first Python-level rule is labeled \emph{P1}.
\subsubsection{General Python Rules}

Due to its wide usage and acceptance within the Python community, we base our Python-level rules primarily on those defined and validated by \textit{PyLint}. However, we exclude style-related issues — which do not affect code correctness — as well as runtime errors such as undefined variables, which would prevent the code from executing properly regardless of any linting effort. Instead, we focus on code issues categorized as warnings and refactoring opportunities, as they are more relevant to maintainability and can help uncover potential logical flaws or unintended behaviors.

We selected some of the most frequently violated rules observed in machine learning notebooks \cite{9474395, 10356405}, while excluding those that are not relevant in this context. For example, rules like \texttt{pointless-statement}—which detect statements with no effect—are not appropriate for notebooks, where placing a variable or expression at the end of a cell to display its output is a common and intentional practice.
The selected rules and their descriptions are presented in Table~\ref{tab:python-lint-rules}. Threshold limits used for each rule match those defined by \textit{Pylint}, but they can be easily adjusted according to specific requirements.

\begin{table}[h]
  \centering
  \small
  \caption{General Python Linting Rules Implemented in Vespucci}
  \begin{tabular}{|l|p{5cm}|p{5cm}|}
    \hline
    \textbf{Rule Name} & \textbf{Description} & \textbf{Motivation} \\
    \hline
    \emph{P1} - Unused variables & Identifies variables that are assigned but never used. & Helps remove dead code and clarify the intent. \\
    \hline
    \emph{P2} - Variables reassignments & Detects repeated use of the same variable name after initial assignment. & Prevents confusion and potential bugs in notebooks. \\
    \hline
    \emph{P3} - Variables naming & Enforces a minimum variable name length (more than 3 characters, except for \gls{ml} conventions like \texttt{X}, \texttt{y}). & Promotes readability and avoids ambiguous identifiers. \\
    \hline
    \emph{P4} - Too many parameters & Detects functions with more than 5 parameters. & Encourages simpler, modular function design. \\
    \hline
    \emph{P5} - Consider using from import & Recommends using \texttt{from module import x} instead of importing the full module. & Improves clarity and avoid namespace pollution  \\
    \hline
    \emph{P6} - Unused import & Identifies unused imports with an exception for the star import (\texttt{from module import *}) where we can not statically determine which symbols are actually used in the code.  & Improves clarity. \\
    \hline
    \emph{P7} - Reimported & Identifies redundant imports of the same module. & Reduces code clutter and redundancy. \\
    \hline
    \emph{P8} - Too many local & Detects functions that declare more than 11 local variables. & Indicates high complexity. \\
    \hline
    \emph{P9} - Global variables in function & Warns against the use of global variables within functions. & Prevents hidden dependencies and side effects. \\
    \hline
  \end{tabular}
  \label{tab:python-lint-rules}
\end{table}

\subsubsection{ Notebook-specific Rules}
\noindent
The following linting rules have been defined specifically for notebook-level quality, sourced from literature \cite{QuarantaA2022,10.1371/journal.pcbi.1007007}. These rules address common issues in code organization, naming, and execution behavior that affect readability, reproducibility, and maintainability. Naming conventions and threshold limits used for each rule match those defined by \textit{Pynblint}, and can be easily modified if needed. Unlike \textit{Pynblint}, which includes checks requiring external context (e.g., presence of git repositories), we focus exclusively on rules verifiable using only the notebook file itself.
Selected rules and their descriptions are presented in Table~\ref{tab:notebook-lint-rules}.

\begin{table}[h]
  \centering
  \small
  \caption{Notebook-Level Linting Rules Implemented in Vespucci}
  \begin{tabular}{|l|p{5cm}|p{5cm}|}
    \hline
    \textbf{Rule Name} & \textbf{Description} & \textbf{Motivation} \\
    \hline
    \emph{N1} - Notebook naming & Enforces the use of filenames longer than 2 characters (\emph{N1.1}) with only alphanumeric characters, hyphens, or underscores (\emph{N1.2}). Detect default names containing \texttt{Untitled} (\emph{N1.3}). & Improves clarity, navigation, and compatibility across systems. Prevents confusion during sharing or version control. \\ 
    \hline
    \emph{N2} - Version control & Requires displaying library versioning information with \texttt{package.\_\_version\_\_} or using \texttt{watermark}\footnote{\url{https://github.com/rasbt/watermark}}. & Ensures reproducibility and helps diagnose variances across executions or environments. \\
    \hline
    \emph{N3} - Imports at the top & Enforces grouping all import statements in the first code cell. & Enhances readability and makes dependencies explicit for reproducibility. \\
    \hline
    \emph{N4} - Long code cells & Detect cells exceeding 30 lines of code. & Encourages modular design, improves readability. \\
    \hline
    \emph{N5} - Empty code cells & Detects code cells that contain no code. & Avoids unnecessary visual clutter and potential confusion. \\
    \hline
    \emph{N6} - Missing text cells & Checks for absence of markdown cells providing context or explanation. & Promotes documentation and guides readers through the workflow. \\
    \hline
    \emph{N7} - Notebook too long & Detect notebooks with more than 50 code cells. & Notebooks should remain reasonably short and focused. \\
    \hline
    \emph{N8} - Non-linear execution & Detects out-of-order execution based on cell execution counts (if present). & Ensuring a linear execution order helps maintain a clear, predictable
    workflow and supports reliable re-execution of the notebook from top to bottom. \\
    \hline
  \end{tabular}
  \label{tab:notebook-lint-rules}
\end{table}

\subsubsection{Machine Learning-specific Rules}
Machine learning-specific rules are derived from literature \cite{9609188,10.1145/3522664.3528620} and focus mostly on correct usage patterns of ML-related libraries, particularly \texttt{pandas} (version 2.2) and \texttt{scikit-learn} (version 1.6.1). Our framework is extensible, allowing easy integration of additional libraries and checks.
Selected rules and their descriptions are presented in Table~\ref{tab:ml-lint-rules}.

\begin{table}[h]
  \centering
  \small
  \caption{ML-Specific Linting Rules Implemented in Vespucci}
  \begin{tabular}{|l|p{5cm}|p{5cm}|}
    \hline
    \textbf{Rule Name} & \textbf{Description} & \textbf{Motivation} \\
    \hline
    \emph{M1} - Uncontrolled randomness & Detects missing random seed parameters in ML-related functions (e.g., train/test split, model initialization). & Ensures reproducibility of results and consistency across runs. \\
    \hline
    \emph{M2} - In-place API misuse & Warns when functions like \texttt{dropna()} are used without assigning the result. & Prevents logic errors due to misunderstanding between in-place modification and returning copies. \\
    \hline
    \emph{M3} - Implicit hyperparameters & Detects ML model instantiations where key hyperparameters are not explicitly set. & Enhances transparency, reproducibility, and adaptability to future library changes. \\
    \hline
    \emph{M4} - Columns and dtypes not set & Detects when columns (\emph{M4.1}) and datatypes (\emph{M4.2}) are not explicitly specified during data loading. & Improves data integrity, parsing performance, and avoids unintended type inference. \\
    \hline
    \emph{M5} - Merge without explicit parameters & Detects DataFrame merges where parameters \texttt{on}, \texttt{how} (\emph{M5.1}), or \texttt{validate} (\emph{M5.2}) are not specified. & Reduces risk of ambiguous or incorrect merges and improves code clarity. \\
    \hline
  \end{tabular}
  \label{tab:ml-lint-rules}
\end{table}

\section{Exploratory Analysis on Real-World Python Notebooks}
To assess the practical relevance of our linter, we applied it to a representative sample of 5,000 real-world Jupyter notebooks from the Kaggle platform. This empirical evaluation aims to answer whether real notebooks exhibit rule violations detectable by our tool and, if so, which types of violations are most common, and what do they reveal about prevalent coding and notebook practices.

By aggregating and analyzing the rule violations detected across the dataset, we were able to uncover usage patterns and recurring issues—from common structural inconsistencies (e.g., misplaced imports) to ML-specific issues such as lack of control over randomness or missing parameters in data loading.

As we will see, certain rules are violated in a majority of notebooks, sometimes repeatedly within the same notebook, while others are rarely or never triggered. This analysis validates the utility of our tool and also offers insights into current practices in notebook development and potential areas for improvement.

\subsection{Dataset}
\label{sec:dataset}
Our empirical analysis is based on the open-source \textbf{KGTorrent} dataset \cite{9463068}, which contains 248,761 Python notebooks collected from Kaggle between November 2015 and October 2020.
Each notebook is named following the pattern \texttt{<kaggle\_username>+<notebook\_name>} and is organized alphabetically.
For computational feasibility, we randomly selected and analysed 5,000 notebooks from the dataset, without applying any thematic or quality-based filtering.

\subsection{Experimental Setup}
We processed the selected notebooks using our analysis tool, which generates a report for each notebook, listing the violated rules. We then aggregated these individual reports into a unified dataset where each entry corresponds to a rule violation, identified by the notebook name and the rule violated.
\noindent
To analyze the aggregated results, we grouped the data by rules and computed the following statistics: the total number of violations per rule (num\_violations), the number of distinct notebooks in which each rule was violated (num\_notebooks), and the percentage of our analysed notebooks affected by each rule (pct\_notebooks). Additionally, we calculated the average number of violations per affected notebook (violations\_per\_affected\_nb).
\noindent
Each rule was also mapped to the corresponding level (\textit{Python}, \textit{Notebook}, or \textit{ML}). Finally, we organized and sorted the results based on the frequency of violations, and applied a visual style to highlight the most frequent issues.

\subsection{Analysis}

\begin{figure}[ht]
  \centering
  \includegraphics[width=\textwidth]{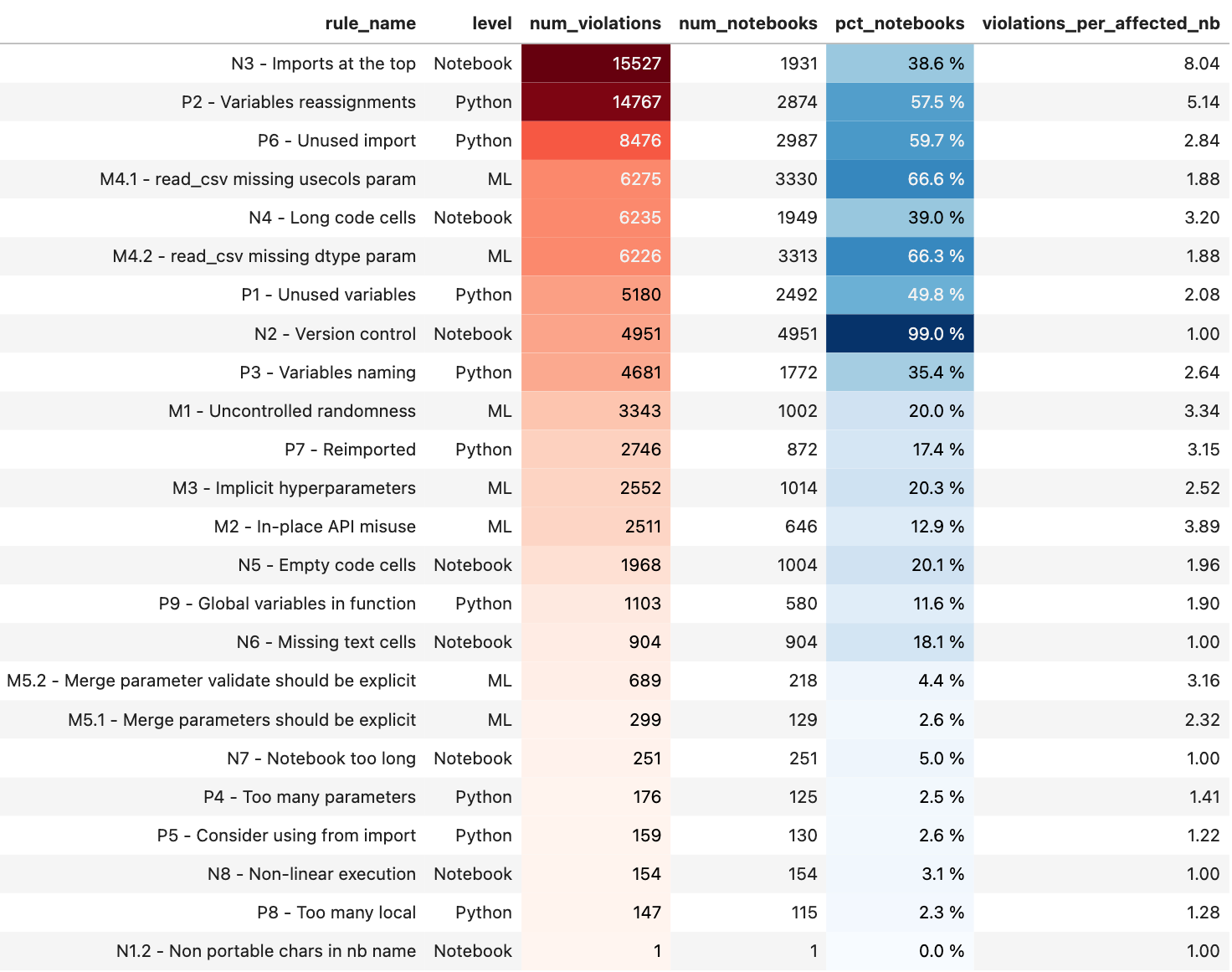}
  \caption{Frequency and distribution of rule violations detected across 5,000 notebooks.}
  \label{fig:nb_violations}
\end{figure}

Figure \ref{fig:nb_violations} displays the rules that we found have at least one violation on our dataset of 5,000 notebooks. 
During the experiment, we conducted a manual verification to assess the accuracy of the detected violations. We selected 15 notebooks and reviewed the violated rules. This inspection revealed that most rules were correctly triggered. However, we identified consistent false positives for the rule \textit{Unused Variable}. These errors are due to technical limitations in our current Python parsing mechanism. This limitation will be addressed in future work to improve detection precision.
The analysis of the 5,000 notebooks from our subsample reveal several patterns in the rule violations detected by our tool. 
\begin{description}
\item[Frequent multiple violations per notebook.] Some rules are frequently violated and tend to appear multiple times within the same notebook. The rule \textit{N3 - Imports at the top} was violated 15,527 times across 1931 notebooks—38.6\% of the sample—with an average of 8.04 violations per affected notebook. Similarly, the rule \textit{P2 - Variable reassignments} appear in 2874 notebooks (57.5\%) with a total of 14,767 violations (5.14 per notebook). 
These findings could suggest that once a certain practice is adopted in a notebook, it tends to be consistently repeated throughout. 
\item [Widespread but single-instance violations.] Some rules are structurally limited to a single occurrence per notebook. This is the case for \textit{N2 - Version control}, which can only be violated once per notebook. As we can see, it is violated in 4951 notebooks—99.0\% of the dataset. This high prevalence may reflect different possibilities: our rule definition might be too narrow, and alternative detection strategies could be more appropriate, environment-level indicators (e.g., \texttt{requirements.txt}, Git) should be included in the analysis, or a widespread lack of version control usage on the studied notebooks (which is the least likely). This contrasts with rules like \textit{N3 - Imports at the top} , which, despite a higher absolute number of violations, only affects 38.6\% of notebooks.
\item[Lack of textual documentation.] The rule \textit{N6 - Missing text cells} is violated in 904 notebooks, representing 18.1\% of our sample. Each violation occurs once per notebook by design. This could mean those notebooks might be used for rapid prototyping or personal experimentation rather than structured communication or result sharing.
\item[Machine Learning rules.] Rules specific to \gls{ml} highlight some common issues. The omission of \texttt{usecols} and \texttt{dtype} parameters in \texttt{read\_csv} calls is very common, with 6275 and 6226 violations respectively, occurring in over 65\% of notebooks. Reproducibility-related rules like \textit{M1 - Uncontrolled randomness} (3343 violations in 20\% of notebooks) and \textit{M3 - Implicit hyperparameters} (2552 violations in 20.3\%) are also prevalent. This could indicate a lack of attention to experimental reproducibility and parameter tracking. 
\item[Rarely violated rules] Some rules are rarely triggered. For instance, \textit{P4 - Too many parameters}, \textit{P8 - Too many locals}, and \textit{P5 - Consider using from-import} each appear in fewer than 3\% of notebooks. 
For P4 and P8, this might reflect moderate function complexity across the dataset or a tendency toward simple, single-purpose functions. It is also possible that the thresholds defined for these rules are too high or that very few functions were defined. 
\item[Marginal or non-appearing rules.] Finally, we can see the almost absence of violations of the rules \textit{N1 - Notebook naming} with a single violation. As described in Section~\ref{sec:dataset}, all notebook filenames were normalized in the original dataset, which prevents this rule from being triggered.
\end{description}

\subsection{Threats To Validity}

Several threats to validity should be considered when interpreting our findings.
\noindent
During manual inspection, we identified false positives associated with the \textit{Unused Variable} rule. These are due to current limitations in our tool, which will be addressed in future work. For all other rules, however, no systematic false positives were observed.
\noindent
Also, our static analysis tool is intentionally designed to favor false negatives over false positives. This approach can help avoid overreporting violations, but it also implies that our results may underestimate the true number of violations present in the notebooks.
\noindent
Finally, our analysis is based on a relatively small sample of 5270 notebooks, all sourced from the Kaggle platform. This limits the generalizability of our findings, as these notebooks may not be representative of broader notebook usage patterns across other platforms.

\section{Conclusion and Future Work}
In this paper, we introduced \textit{Vespucci Linter}, an analysis tool designed to detect bad practices in machine learning notebooks across three levels: general Python code, notebook structure, and \gls{ml}-specific workflows. 
Unlike existing tools that focus on a single level, our tool leverages metamodels to provide richer and more contextualized analysis.
We implemented 22 linting rules based on state-of-the-art literature and validated their effectiveness through an exploratory analysis of 5,000 real-world notebooks collected from the Kaggle platform. 
The results highlight widespread violations on the three levels which confirms the relevance of our multi-level approach.

Diverse directions remain to be explored in future work. First, we plan to integrate our tool more directly into notebook environments, such as JupyterLab or VSCode to provide real-time overlays or inline annotations. This would allow developers to receive immediate feedback as they write or modify notebooks.
Additionally, we aim to develop more semantically rich linting rules that take full advantage of our tool’s capabilities. Future rules could focus on deeper semantic aspects such as pipeline structure validation or data leakage detection—analyses that are difficult to implement with traditional linters.
Finally, a more extensive empirical study is needed to explore the early results introduced in the exploratory analysis.

\begin{acknowledgments}
  This work was supported by the French National Research Agency (ANR) under project ANR-AAPG2024, PROFIL.
\end{acknowledgments}

  

\bibliography{references}

\end{document}